\documentclass{svproc}

\usepackage{url}

\usepackage{graphicx}
\usepackage{hyperref}
\usepackage{enumerate}
\usepackage{bbding} 
\usepackage{xcolor}
\usepackage{soul}
\sethlcolor{black}
\makeatletter
\newif\if@blind
\@blindtrue 
\if@blind \sethlcolor{black}\else
   
\fi

\usepackage{bbding} 

\begin{document}

\newcommand{\ecotwin}{Ecotwin }
\newcommand{\www}{www.ecotwin.se}

\mainmatter              
\title{
AI Tool for Exploring How Economic Activities Impact  Local Ecosystems
}
%
\titlerunning{AI Tool for Exploring Ecosystems}  
\author{Claes Strannegård\inst{1}{}\Envelope
\and Niklas Engsner\inst{2}
\and Rasmus Lindgren\inst{2}
\and Simon Olsson\inst{2}
\and John Endler\inst{3}}
\authorrunning{Claes Strannegård et al.} 
\tocauthor{Claes Strannegård, Niklas Engsner, Rasmus Lindgren, Simon Olsson, and John Endler}
\institute{
Applied Information Technology, University of Gothenburg, Gothenburg, Sweden\\
\email{claes.strannegard@gu.se}
\and
Computer Science and Engineering, Chalmers University of Technology, Gothenburg, Sweden
\and
School of Life and Environmental Sciences, Deakin University, Waurn Ponds, Australia
}

\maketitle              

\begin{abstract}
We present an AI-based ecosystem simulator that uses three-dimensional models of the terrain and animal models controlled by deep reinforcement learning. The simulations take place in a game engine environment, which enables continuous visual observation of the ecosystem model. 
The terrain models are generated from geographic data with altitudes and land cover type. The animal models combine three-dimensional conformation models with animation schemes and decision-making mechanisms trained with deep reinforcement learning in increasingly complex environments (curriculum learning).  
We show how AI tools of this kind can be used for modeling the development of specific ecosystems with and without different forms of economic activities. 
In particular, we show how they might be used for modeling local biodiversity effects of 
land cover change,
exploitation of natural resources,
pollution,
invasive species, and 
climate change. 

\keywords{
economic activities,
ecosystem simulator,
3D terrain model, 
agent-based model,
deep reinforcement learning}
\end{abstract}
\section{Introduction}

Biodiversity is fundamental to human well-being and economic prosperity, yet
 human economic activities continue to reduce biodiversity on a massive scale   \cite{dasgupta2021economics,cardinale2012biodiversity}.
The main drivers of anthropogenic biodiversity loss in terrestrial ecosystems are (in decreasing order of importance)
land use change,
direct exploitation, 
pollution,
invasive species, and
climate change 
\cite{jaureguiberry2022direct}.
To be able to combine economic development with ecological sustainability 
in the long run, the sequence of human damage to ecosystems needs to be halted and reversed  \cite{kunming2022montreal}. 
A central part of that challenge is understanding how different forms of human economic activity might impact specific ecosystems. 
With such an understanding, disastrous decisions can be avoided already at the planning stage and more sustainable alternatives can be chosen. 

Ecosystem simulators can play an enabling role in such decision-making provided they are good enough at making predictions about ecosystem development. In particular, they must not omit any of the main factors that influence ecosystem development. 
One such factor is the terrain, as determined by the altitudes and land cover types of an area. In fact, the terrain defines the three-dimensional habitat of animals with its food sources, water bodies, hiding places, and shelters from extreme temperatures. It also brings relative advantages and disadvantages to predators and prey by influencing their perception and mobility. 


Despite the fundamental role played by the terrain in ecosystems, ecosystem simulators typically do not model it at all, or model it at very low level of detail.
The purpose of this paper is to fill this gap by presenting an ecosystem simulator that includes detailed models of the terrain based on geographic data. We combine this with animal models, whose behavior is generated by deep reinforcement learning.

Sections \ref{sec:ecosystem_models} and \ref{sec:Challenges} give an overview of ecosystem models and  some of their limitations.
In Section \ref{sec:Ecotwin} we introduce our simulator Ecotwin. Then we go into more depth with its terrain model in Section \ref{sec:terrain_model} and its plant and animal models in Section \ref{sec:animal_model}.
In Sections \ref{sec:baselines} and \ref{sec:studies}, we illustrate how \ecotwin can be used to model ecosystems with and without various forms of economic activity.
Sections \ref{sec:discussion} and \ref{sec:conclusion}, finally, contain a discussion and a conclusion.


\section{Ecosystem models}
\label{sec:ecosystem_models}

Ecosystem data repositories are sometimes viewed as  ecosystem models. Examples include the classic Hudson's Bay dataset, with population data for snow-shoe hares and lynx \cite{elton1942ten}, and the dataset from the tropical island of Moorea, with data ranging from weather to genomes \cite{cressey2015tropical}. Data repositories alone are clearly not enough for making predictions about ecosystem development, however.

Turning to ecosystem models that might be used for prediction purposes, we have population-based models (PBM), which represent each type of organism numerically at the aggregated level \cite{royama2012analytical}. For instance, one may use a variable for the number of hares in a certain terrestrial area or the tonnes of cod in a certain marine area.
Classic examples of PBMs include exponential growth models \cite{malthus1986essay},
logistic growth models \cite{verhulst1838notice}, Lotka-Volterra models \cite{lotka1925elements}, and Leslie models \cite{leslie1945use}. Note that these models do not take terrain data into account. 
%
There are also PBMs that include space representations. Some of them apply a grid to a map and introduce population variables for each cell of the grid, for instance, the number of hares in a certain 1 km$^2$ square on a map or the tonnes of cod in a certain 1 km$^2$ square on a sea chart. An example of such a model is the Ecopath (with Ecosim and Ecospace) simulator for marine ecosystems \cite{christensen2004ecopath}. These models require specifications at the aggregated level of how the population quantities of each cell depend on other population quantities of the same cell and its neighboring cells. 

Another type of ecosystem model is the individual-based model (IBM) \cite{deangelis2014individual}, also called agent-based model (ABM), which models each organism separately. 
The IBMs include manually programmed models, probabilistic models, and AI-generated models.

Manually programmed IBMs are used in simplistic animal models such as Braitenberg vehicles \cite{braitenberg1986vehicles}. They are also ubiquitous in game AI, where models of animals and imaginary creatures are controlled by finite state machines, sometimes with added noise \cite{bourg2004ai}. A classic example are the ghosts of Pac-Man \cite{pacman}. Manually programming animal models that are able to handle all situations that might arise in complex ecosystems, seems to be extremely challenging. 

Examples of probabilistic models include those of the Neutral Theory of Biodiversity \cite{hubbell2001unified}, 
sometimes combined with models based on cellular automata \cite{missa2016understanding}. 
While those particular models have great theoretical interest, they are not intended for modeling concrete ecosystems. In fact, they use rudimentary models of animal and plant behavior, while the terrain is essentially not modelled at all.

There are several kinds of AI-controlled animal models.  
Some of them rely on evolutionary algorithms \cite{coello2007evolutionary}, where one idea is to evolve parameters of neural networks that control behavior. This has been tried in the context of video games \cite{salimans2017evolution} and for modeling rats in mazes, using recurrent neural networks \cite{zou2021neuroevolution}.
Another kind of AI-controlled animal model relies on reinforcement learning (RL)
\cite{sutton2018reinforcement}, in particular deep reinforcement learning (DRL) \cite{arulkumaran2017deep}, which is based on deep learning \cite{goodfellow2016deep}. DRL has been highly successful at playing video games, reaching super-human performance on all classic Atari 2600 games \cite{badia2020agent57}. It has also been used for playing go and chess at super-human level \cite{silver2018general} and for controlling helicopters and drones that perform acrobatic tricks \cite{abbeel2006application}. 
Moreover, DRL has been used for training predator and prey models in maze-like worlds \cite{yamada2020evolution}, even with millions of agents \cite{yang2017study}. It has also been used for training animal models to survive in toy worlds with herbivores, predators and apex predators \cite{sunehag2019reinforcement}. 

\section{Challenges and limitations}
\label{sec:Challenges}

We know beforehand that predicting the exact development of ecosystems -- pristine or subject to human interventions -- is a formidable challenge. For one thing, our knowledge about exact animal behavior is limited \cite{parrott2010measuring}. 
Still, what we know might be enough for making useful approximations regarding the population dynamics of a few species that are connected to each other via the food web. 
Such approximations can be useful for decision-makers seeking to understand the consequences of different policies or business decisions.

A common problem is that the ecosystem model relies on manually programmed animal models. In fact, this is overwhelmingly hard, except in highly simplified cases. Here AI and machine learning can come to the rescue, however.
Another common problem is that the ecosystem model does not model the terrain at all (like Lotka-Volterra), or uses a very coarse spatial model (like Ecopath with Ecospace). In fact, this severely limits the ability of the model to predict the development of pristine ecosystems as well as ecosystems affected by economic activities.

Economic activities such as agriculture, forestry, and construction typically lead to radical changes to the land cover type. For instance, natural terrain might be converted into cultivated fields, planted forests, pastures, buildings, dams, quarries, airports, roads, or railways. 
With a terrain model built from geographic data, one may explore the consequences of such changes by simply changing the land cover type and running the resulting model on the simulator. 

 

\section{Ecotwin}
\label{sec:Ecotwin} 

In this section, we give a brief overview of the ecosystem simulator \ecotwin \cite{strannegaard2022ecosystem,strannegard2021ecosystem}. More details follow in the subsequent sections.
As far as we are aware, \ecotwin is among the first ecosystem models based on game engines and AI. It combines high-resolution terrain models with animal models based on deep reinforcement learning. 
%
\ecotwin is built on top of the game engine Unity, which offers a rich environment with graphical representations, animations, and a physics engine \cite{unity}. We use ready-made Unity models of animals and plants. The animal models look much like real animals, but they have no decision-making mechanism or ``brain''. We fill this gap by equipping the animal models with a generic decision-making mechanism based on deep reinforcement learning. 
\ecotwin is open source and available at 
\cite{ecotwin}, where several simulation videos can also be found.

\section{Terrain model}
\label{sec:terrain_model}  

\begin{figure}[ht]
  \centering
    \includegraphics[width=0.6\linewidth]{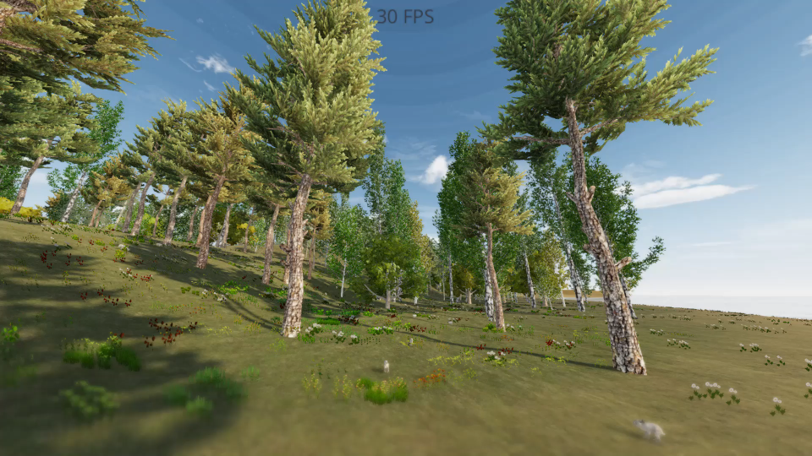}    
    \caption{View of the Unity model of the island with barely visible hares. 
    }
   \label{fig:grass}
\end{figure}

The terrain models of \ecotwin are generated by Unity from geographic data with altitudes and land cover type. Such data are available for large portions of the world. In our case studies, we used terrain data from open sources at the Swedish University of Agricultural Sciences \cite{SLU} and the Swedish Environmental Protection Agency \cite{landcoverdata}. 
For the sake of illustration, we focused on Lilla Amundön, a small island on the Swedish west coast, which fits into a $1\times 1$ km square. It only took Unity a second to generate a 3D-model of the island with different land cover types. Then, we automatically placed tree models on the surface of the model in a random fashion, with probabilities conditioned on the land cover type. For example, pine tree models were put on the land cover type pine forest. We also distributed 1000 grass patches and 1000 dandelions on random locations in fields and forests, with a relatively high probability in the land cover type field and less in the forest. 
These plants were used, respectively, as sources of glucose and hydration for the hare models of the island, which will soon be introduced.
Fig. \ref{fig:grass} shows a view of the resulting Unity model. 

\section{Animal model}
\label{sec:animal_model} 

\begin{figure}[ht]
  \centering
    \includegraphics[width=0.6\linewidth]{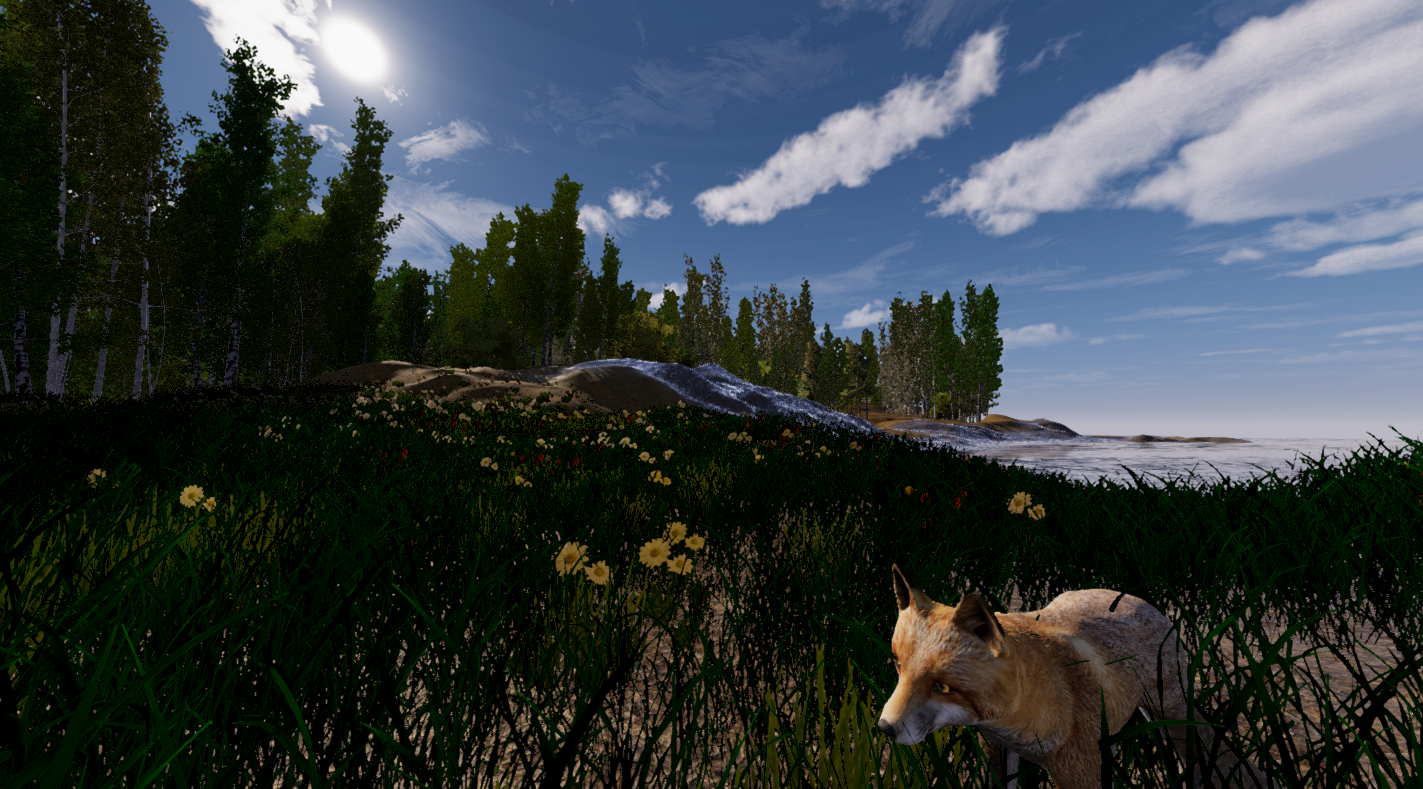}    
    \caption{View of the Unity model of the island with a fox. 
    }
   \label{fig:fox}
\end{figure}

We use ready-made Unity models of two animal species: European hare (\textit{Lepus europaeus}) and red fox (\textit{Vulpes vulpes}). 
Fig. \ref{fig:fox} shows a view of the Unity model with a fox. 
As mentioned, the ready-made animal models, including the hare and fox models,  come without decision-making mechanisms. In this section we will show how to construct such mechanisms based on neural networks. 
For simplicity we use one neural network for the decision-making of all hares and another for all foxes. These networks make different decisions depending on the input to each individual hare or fox. 
Let us describe these decision-making mechanisms in more detail. 

\subsection{Perception}

The perception model is divided into exteroception, interoception, and proprioception.  At each step, the perception signals are fed into a neural network that computes the actions of the animal model.

\subsubsection*{Exteroception}

\begin{figure}[ht]
  \centering
    \includegraphics[width=0.4\linewidth]{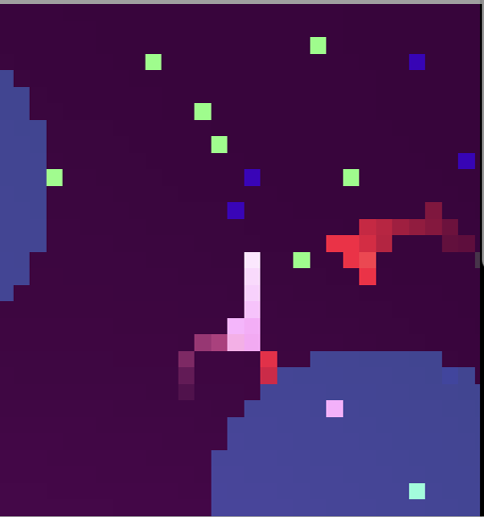}
    \caption{Image input to a hare model. The white pixel at the center of the image marks the hare's own position, while the light purple pixels mark its last few positions. 
    The red pixels represent two other hares and their last few positions. 
    There are also pixels representing grass patches (green), dandelions (ultramarine), field  (purple) and forest (lavender). 
   }
    \label{fig:mindmap}
\end{figure}

This type of perception includes a $31\times 31$ pixel image, covering a $31\times 31$ meter area centered around the agent. Intuitively, that image summarizes sensory data from several senses and a short-term memory. The agent is always at the center of the image and north is always upward in the image. 
The images for foxes and hares contain slightly different information.  
Each image consists of several channels that provide information about each $1\times 1$ meter square inside the $31\times 31$ meter area. For both animals, the image contains the following information:
\begin{itemize}
\item altitude
\item land cover type, e.g., forest, field, rock, sea
\item presence of grass patches
\item presence of dandelions
\item presence of obstacles, e.g., trees
\item presence of hares and foxes
\end{itemize}
An example of such an image is shown in Fig. \ref{fig:mindmap}. 
In addition, exteroception includes two smell vectors, indicating the aggregated smell of hares and foxes, respectively, within a 100 meter radius. 



\subsubsection*{Interoception}

The  following interoceptive signals are also part of the input to the neural network:
\begin{itemize}
\item glucose level
\item hydration level
\item utility level (defined below)
\end{itemize}

\subsubsection*{Proprioception}

Finally, the following signals are part of the input:
\begin{itemize}
\item the present orientation of the self
\item the last few positions of the self (the agent in the middle)
\end{itemize}

\subsection{Actions}
Both animals have four actions: 
\begin{itemize}
\item stand still (and eat)
\item move forward
\item turn left 15 degrees
\item turn right 15 degrees. 
\end{itemize}
Animals can move forward up and down slopes with a maximum inclination of 45 degrees. 
Newborn animals stand still for about 100 time steps after they are born. Foxes require about 50 time steps to eat a hare. 
 

\subsection{Metabolism}

Hares and foxes need two resources to stay alive: glucose and hydration. The resource range of the hare is $[0, 5]$ and of the fox $[0, 8]$. When the level is at the max value, consuming more will not result in higher levels of resources. Both the hares and the foxes have the initial value $2$ for both resources.

\begin{itemize}
\item Resource gain: Unless they are full, hares get a glucose addition of +1 for eating a patch of grass and an addition of +1 hydration for eating a (succulent) dandelion. 
Unless they are full, the foxes get 75\% of the resources of the hares that they eat. 

\item Resource loss: The hare and fox models have a basic metabolic cost of 0.001 per step for each resource.
Both models have additional costs for moving uphill, depending on the slope (up to 0.001 per step). 
Animal models that walk into trees will lose 0.1 of each resource (from ``bleeding''). 
\end{itemize}
These resource changes will affect the utility of the animal model and thus the reward, according to the definitions (\ref{utility}) and (\ref{reward}).

\subsection{Reproduction}

For simplicity, a very basic haploid reproduction model was used for the hares and foxes. Hares and foxes that have reached maturity age and have sufficiently high levels of resources (above 3.5 for hares and 5.5 for foxes) have a certain 
probability of reproducing. The reproduction probability is  higher if they have more resources. After a gestion time, an offspring is born and placed next to the parent. The offspring have an initial value of 2 for both glucose and hydration. The same amount of resources are deducted from the parent.



\subsection{Death}

Animal models die if any of the following happens: 
\begin{itemize}
\item their glucose or hydration level reaches 0
\item they reach their max age
\item they are killed by a predator
\item they move into the sea 
\end{itemize}
Grass patches and dandelions that are consumed by hares do not die, but grow out again slowly on the same spot, as determined by a parameter.


\subsection{Reward system}

We introduce a notion of utility, with the intuitive idea that the animals are rational in the sense that they strive to maximize their utility in the long run (and survive and reproduce as side effects). For simplicity we use the same definition of utility $u(t)$ at time $t$ for both hares and foxes:
\begin{equation}\label{utility}
    u(t) = log(1+glucose(t))+log(1+hydration(t))
\end{equation} 

The log function is used to reflect decreasing marginal utility. For example, eating food item number ten might bring less utility than eating food item number two. Also, high glucose levels do not compensate for low hydration levels and vice versa.
We also need a reward signal so that we can train the animals' decision-making mechanism using RL:
\begin{equation}\label{reward}
    reward(t)=u(t)-u(t-1)
\end{equation} 

This reward might be viewed as a marginal utility. 
Note that the utility defined in (\ref{utility}) is not used as a reward signal. This ensures that the agent is not rewarded for all actions, as soon as it is full and unthirsty. For example, the hare models will typically not be rewarded for approaching fox models. 
The reward signal defined in (\ref{reward}) bears similarities to the one used in standard RL applications like Pac-Man \cite{pacman}, where the agent gets reward when eating food pellets, but not otherwise. 


\subsection{Decision-making}

For their decision-making, our animal models use neural networks, a.k.a. \textit{policy networks}, which map sensory states to actions. The networks were trained using DRL, more precisely, the actor-critic algorithm PPO \cite{schulman2017proximal}, with discount rate 0.99 and learning rate 0.0003. 

Curriculum learning was used for the training the ``brains'' of the animal models and different curricula applied to hare and fox agents. 
%
%
%
The hare agents were trained in five stages: 
\begin{enumerate}[(I)]
\item on a simple $50\times50$ meter square-shaped island with plenty of grass and dandelions and no obstacles except for a surrounding sea, which acted as an enclosure
\item on a similar $100\times100$ meter island with obstacles added in the form of trees 
\item on a similar $200\times200$ meter island with altitudes added
\item on the island model with all features present except foxes
\item on the island model with all features present including foxes
\end{enumerate}

The fox agents were trained in four stages:
\begin{enumerate}[(i)]
\item on a simple $50\times50$ meter square-shaped island with plenty of hares and no obstacles except the sea
\item on a similar $100\times100$ meter island with obstacles added in the form of trees 
\item on a similar $200\times200$ meter island with altitudes added
\item on the island model with all features present 
\end{enumerate}

At each training stage, the agents were trained for 100 episodes, with each episode lasting up to 5000 time steps, but ending earlier if the agent died.


\vspace{6mm}
\noindent \textit{Remark:}
Since the decision-making mechanisms of real animals are the product of evolutionary processes, it might seem natural to use evolutionary algorithms 
\cite{salimans2017evolution} for producing  behavioral models. 
However, modeling evolutionary processes, where  nervous systems evolve in tandem with the whole body, seems to be very challenging conceptually and technically. Thus, we opted for another strategy -- DRL, which is the current state-of-the-art in AI. Again, we are aware that nervous systems did not arise as a long process of RL (although animals use RL), but we still opted for DRL because of its versatility and strength. 

\section{Modeling pristine ecosystems}
\label{sec:baselines}
Here we show how \ecotwin can be used for modeling ecosystems without any human interventions. Videos of simulations can be found at \cite{ecotwin}.
\subsection{Colonization}

\begin{figure}[ht]
  \centering
    \includegraphics[width=0.9\linewidth]{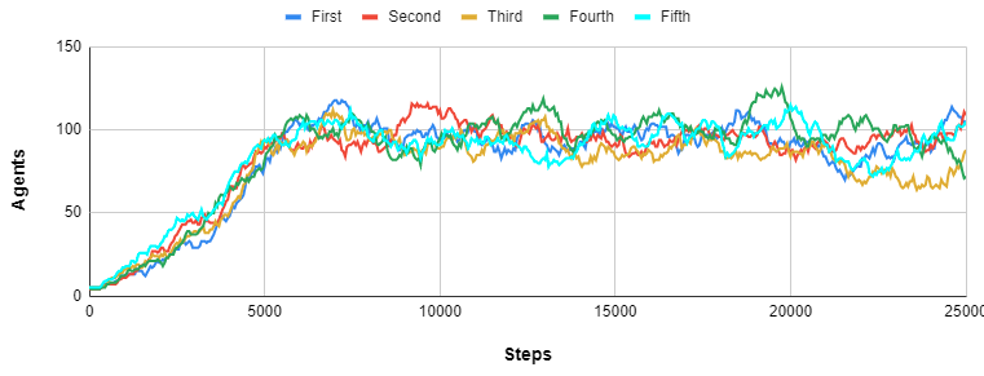}
        \caption{Development of a hare population on the island without any human intervention and without predation.  
        Each curve color represents a different simulation run. In the beginning, five hares reach the shore of the island (perhaps by swimming or crossing the ice) and start colonizing it. In each simulation, we can see that the population grows rapidly in the beginning and then tends to stabilize. There is an element of ``randomness'' in these simulations, which might also be present in real ecosystems. For example, in the simulation represented by the yellow curve, the hare population drops towards the end. This could be due to competition for resources in a relatively inaccessible field, which (temporarily) wipes out the hare population there.}
    \label{fig:baseline}
\end{figure}

Suppose Lilla Amundön had no hares on it initially. Also suppose that five hares reached the island at one point, e.g., by swimming, or running on the ice in winter.
Fig. \ref{fig:baseline} shows the development of the hare population on the island during five different simulation runs. 
In the simulations we can see elements of logistic growth: the population grows rapidly in the beginning and then reaches a relatively stable carrying capacity \cite{verhulst1838notice}. 

\subsection{Predator-prey dynamics} 

\begin{figure}[ht]
  \centering
    \includegraphics[width=0.6\linewidth]{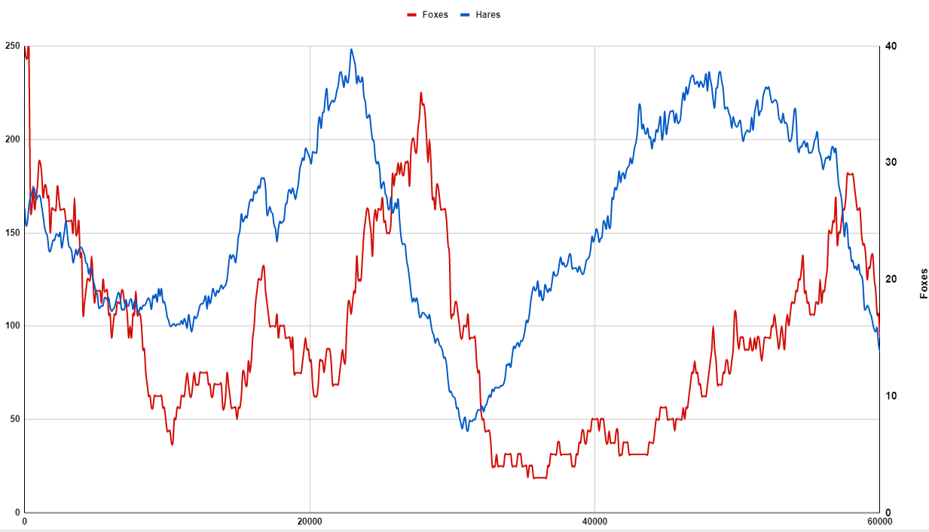}
        \caption{Population curves for hares (blue curve, left scale) and foxes (red curve, right scale). This simulation shows clear elements of the Lotka-Volterra dynamics (coupled predator–prey cycles): 
        (i) when there are relatively few foxes, the hares increase; 
        (ii) when the hares increase, the foxes will do the same with a certain time delay; 
        (iii) when the foxes increase they will kill many hares, so the hares will decrease; and 
        (iv) when the the hares decrease, so will the foxes eventually, since they run out of food. 
        The simulation also shows clear deviations from the Lotka-Volterra model's perfectly regular oscillations, possibly due to randomness and the impact of terrain properties.}
    \label{fig:fox_LV}
\end{figure}

The effect of introducing foxes on the island was studied in three simulations. 
The result of the first one is shown in Fig. \ref{fig:fox_LV}. In the simulation the hare population does not converge at a carrying capacity, as in the baseline case with hares only. Instead we get fluctuating population curves for both hares and foxes. Such fluctuations has been observed in nature, e.g., in the Hudson's Bay data set \cite{elton1942ten} and they have also been predicted by mathematical models, such as Lotka-Volterra. 
We can also see clear deviations from the Lotka-Volterra's perfectly regular oscillations, however.
This comes as no surprise, since randomness (relating to when and where births happen) and terrain properties are also expected to influence the population development.

\begin{figure}[ht]
  \centering
    \includegraphics[width=0.5\linewidth]{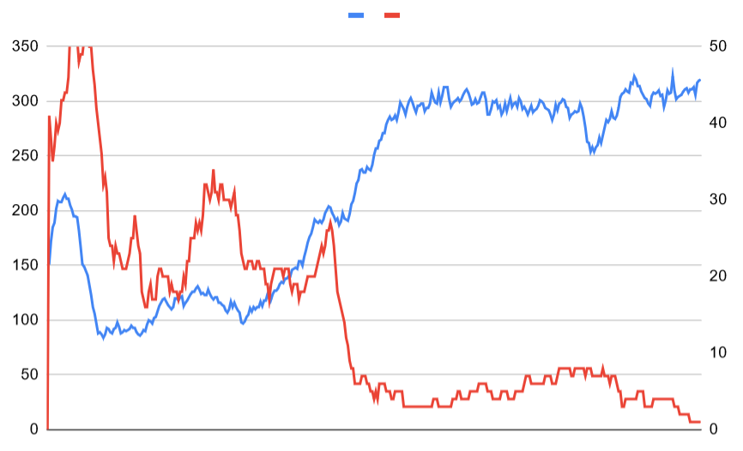}
        \caption{Population curves for hares (blue) and foxes (red). In this simulation, five foxes reach the shores of the island (perhaps by swimming or by crossing the ice) and start colonizing certain parts of the island. We see that the hares survive, but the foxes die out. Some local hare populations might have been wiped out by foxes who then found themselves without food, while other local hare populations (in fox-free areas) managed to survive and eventually take over.}
    \label{fig:hares_survive}
\end{figure}

Fig. \ref{fig:hares_survive} shows the results of the second simulation. Here the hare population first stabilized on the carrying capacity of the island. Then five foxes reach the shores of the island (perhaps by swimming or crossing on the ice in winter) and start colonizing it. 
In this simulation, the foxes eventually died out and the hares survived. One group of foxes moved north, where they could benefit from a relatively dense hare population. Eventually the hares got caught between the sea and the foxes, leading to the extinction of the northern hare population. When there was no food in the north, the foxes turned south and suffered from the fact that they had recently killed the hares in that area.  

This dynamics with an invading predator dying out and the original fauna surviving seems to be more likely in terrain with natural barriers such as heights, bare rock with little food, and water bodies that are difficult to cross. 
On the other hand, the scenario that the invading predators wipe out their prey completely should be more likely to occur in terrain that offers no hiding places for the prey.

\begin{figure}[ht]
  \centering
    \includegraphics[width=0.5\linewidth]{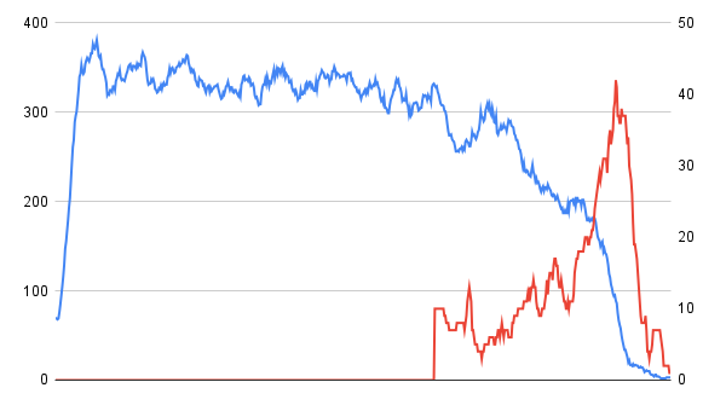}
        \caption{Population curves for hares (blue) and foxes (red).      
        Again the island is invaded by five foxes. In this simulation, the reproduction rate of the foxes was slightly higher. 
        We see that the hares and then the foxes were eventually wiped out.
        }
    \label{fig:fox_no_LV}
\end{figure}

\begin{figure}[ht]
  \centering
    \includegraphics[width=0.5\linewidth]{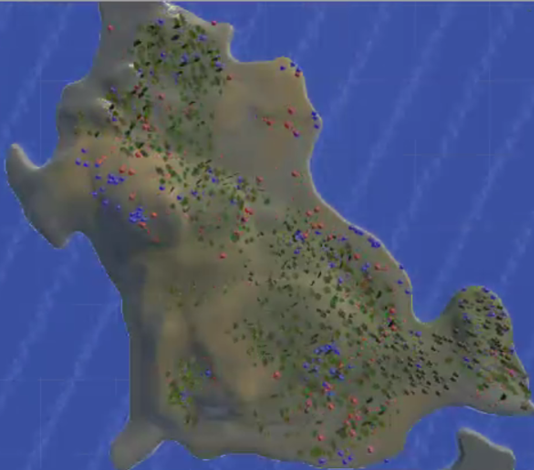}
        \caption{The distribution of hares (blue) and foxes (red), just before the hares died out due to high predation. Then there were no prey for the foxes, so they died out too.}
    \label{fig:many_foxes}
\end{figure}

Fig. \ref{fig:fox_no_LV} shows the result of a third simulation, which used a slightly higher reproduction rate for the foxes. 
This enabled the foxes to reproduce faster and kill the hare populations more efficiently over time. As a result, the hares were wiped out, followed by the foxes. 
Fig. \ref{fig:many_foxes} shows the animal distribution towards the end of this simulation, when there are too many foxes for the hare population to survive. 

These three simulations illustrate that small changes can lead to dramatically different outcomes: both predators and prey survive, only the prey survive, or none of the species survives. 
For that reason not least, it is important to run many simulations with many variations in the context of decision-making. For example, in the end the simulations might suggest that there is a 20\% chance that the hares will be eradicated if foxes were to set foot on the island.


\section{Modeling the impact of economic activities}
\label{sec:studies}

Here we show how \ecotwin can be used for modeling the impact of different forms of human interventions, again on Lilla Amundön. The idea is to make such interventions \textit{in silico} first, so that unintended consequences can be avoided. 


\subsection{Land use change}

\subsubsection{Agriculture}
It is straightforward to change the land cover type, e.g. from forest or field to cultivation. Simulations might reveal  proportional population reduction due to habitat loss. They may also reveal barrier effects as animal models become less motivated or physically prevented from crossing cultivated fields, perhaps reducing accessibility to other feeding grounds, with more far-reaching consequences.

\subsubsection{Forestry}
The land cover type may also be changed to simulate forestry, e.g., from type field to type forest to simulate afforestation or the other way around to simulate logging. The vegetation will follow the land cover type, so the habitats will change, possibly with consequences at the population level. To model the landscape after logging, one may want to introduce a new land cover type, maybe with lower locomotion speed for certain species.

\subsubsection{Urbanization}

\begin{figure}[ht]
  \centering
\includegraphics[width=0.5\linewidth]{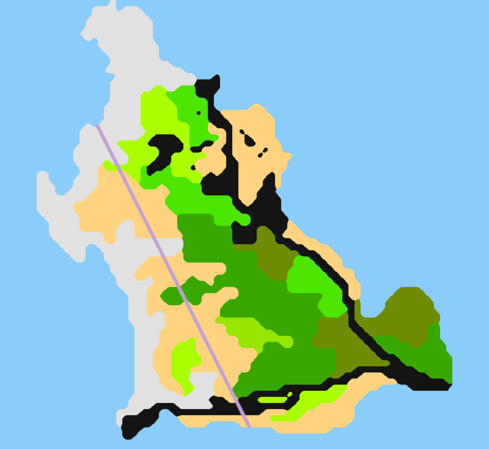}
        \caption{Road stretch. The projected road stretch is shown as a purple straight line. 
        The simulations explore how different levels of traffic intensity might affect the hare population via road kills.}
    \label{fig:road_map}
\end{figure}

\begin{figure}[ht]
  \centering
    \includegraphics[width=0.9\linewidth]{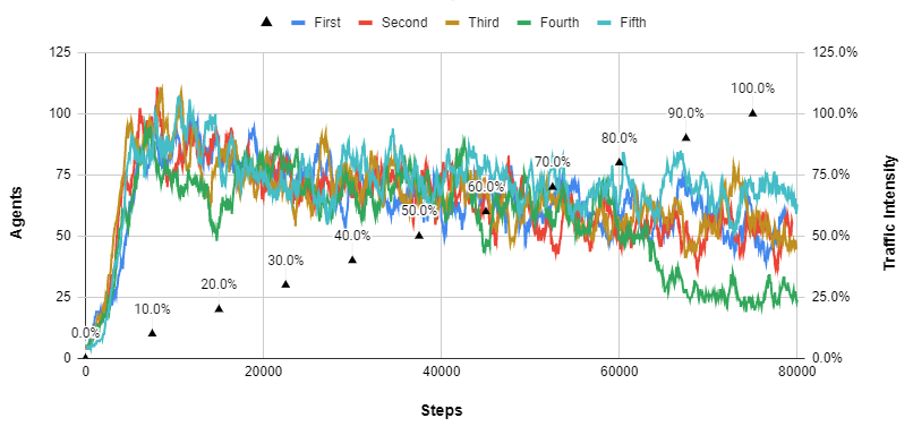}
\caption{Road kills. The traffic intensity on the road stretch was increased gradually and step-wise. The probability for a hare crossing the road of getting killed increased gradually from 0 to 100\%, as indicated by the percentages next to the little black triangles. Each curve color represents a different simulation run. The simulations indicate that the hare population of the island can absorb a 20\% road kill probability without decreasing. In the simulation represented by the green curve, the hare population on the west side of the road went extinct towards the end of the simulation, which explains the dip in that curve.}
    \label{fig:construction}
\end{figure}

The effect of constructing a road on Lilla Amundön, according to the road stretch showed in Fig. \ref{fig:road_map}, was explored in five simulations with steadily increasing traffic intensity during each simulation, resulting in more and more road kills. 
Gradually increasing intensity can be used as a first analytical step for identifying critical levels. As a second step, one may explore the effect of various constant intensity levels. 
The development of the hare population during the simulations is shown in Fig. \ref{fig:construction}. During each simulation, the traffic intensity is not constant, but steadily increasing. The purpose is to get a rough overview of the impact at different intensity levels. These simulations can be followed up with simulations of constant traffic intensity.
It would also  be straight-forward to study the effects of putting fences of both sides of the road that prevent the hares and foxes (or only the foxes) from crossing the road. 




\subsection{Direct exploitation}

\begin{figure}[ht]
  \centering
    \includegraphics[width=0.9\linewidth]{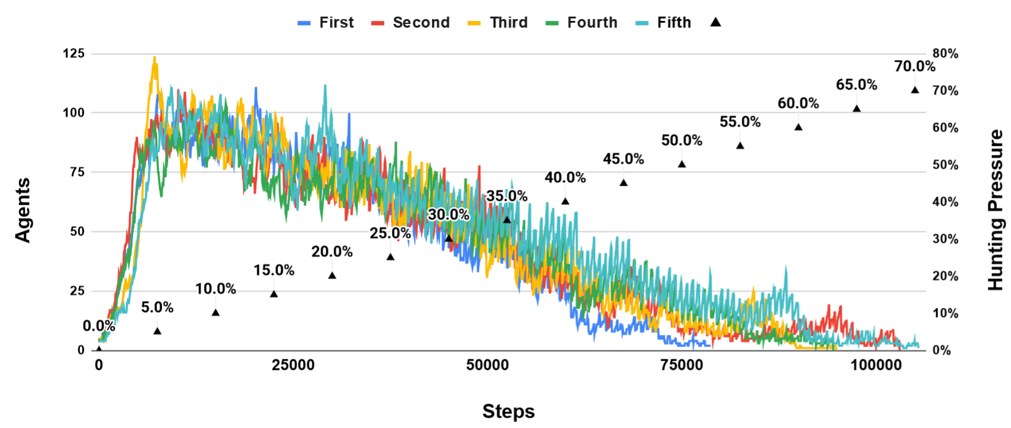}
        \caption{The effect of hunting. The hunting pressure increases gradually and step-wise, as indicated by the percentages next to the little black triangles. These simulations indicate that the population can absorb a 10\% hunting pressure, essentially without decreasing. 
        On the other hand, a 40\% hunting pressure results in a roughly halved hare population. As a follow-up study, one may run several simulations with a constant 10\% hunting quota to establish this conclusion more firmly.}
    \label{fig:hunting}
\end{figure}

The effect of hunting on the island was studied in five simulations with steadily increasing hunting pressure. The results are shown in Fig. \ref{fig:hunting}. 

\subsection{Pollution}
Several consequences of pollution might be studied in the present framework. For example, to study water pollution, one may change the land cover type of certain water bodies from clean water, which is drinkable, to polluted water, which is not. 
Another example is to study the effect of reduced fertility caused by pollution. This can be done in our framework by simply changing the parameter that regulates fertility. This can lead to dramatic effects, as we saw with the fox fertility in Section \ref{sec:baselines}.

\subsection{Invasive species} 
A scenario with an invasive species on Lilla Amundön can be modelled similarly to how the fox was modeled in Section \ref{sec:baselines}, e.g., by using mink models rather than fox models. 
For the record, the American mink (\textit{Neogale vison}) was introduced into the Scandinavian fauna in the 20th century and hares are part of its diet.  

\subsection{Climate change}

\begin{figure}[ht]
  \centering
    \includegraphics[width=0.9\linewidth]{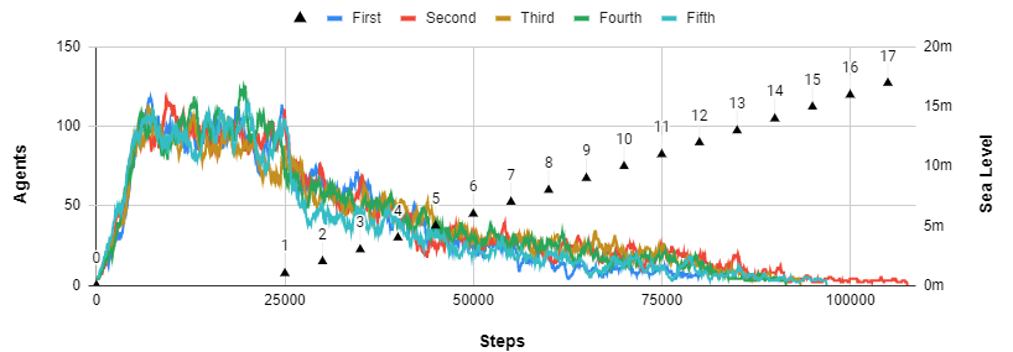}
        \caption{Rising sea levels. The sea level was increased gradually and step-wise, as indicated by the little black triangles. Here we see a clear effect on the hare population already at the 1-2 m level. The population does not go extinct until several meters later, however, since the island also has higher-altitude terrain, where grass and dandelions grow.}
    \label{fig:rising}
\end{figure}

\begin{figure}[ht]
  \centering
    \includegraphics[width=0.5\linewidth]{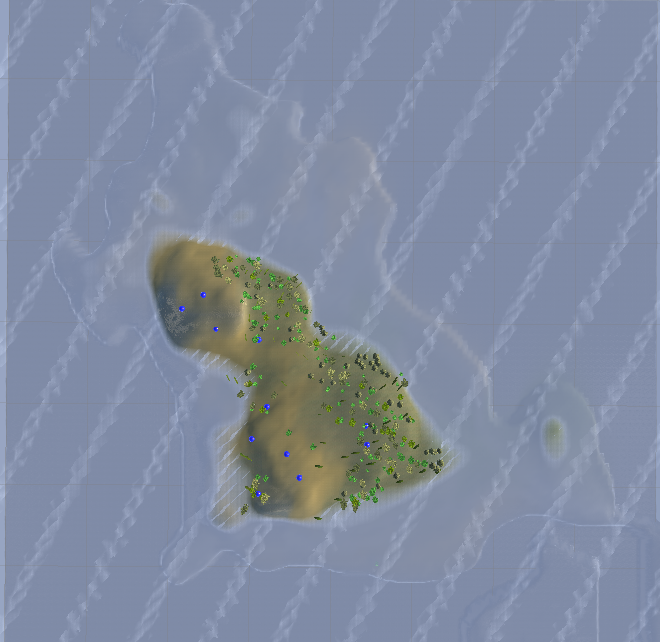}
        \caption{Picture of the island during flooding process. The contours of the original island are barely visible around the present island. 
        The blue dots represent hares. The simulations confirm that the original carrying capacity has been severely reduced. 
        }
    \label{fig:flooded}
\end{figure}

The effect of rising sea levels on the hare population of the island was studied in five simulations. The results are showed in Fig. \ref{fig:rising}. The simulation results indicate dramatic effects already at the 1-2 meter level. Part of this effect is due to habitat loss, given that large portions of the island is low. 
Fig. \ref{fig:flooded} shows the animal distribution of the simulation, when part of the island has been flooded. 

\section{Discussion}
\label{sec:discussion}

\subsection{Validation }
\ecotwin was developed with the following validation criteria in mind: 
\begin{enumerate}

\item To what extent does the model have a realistic look-and-feel in the sense that the animal model look, move, and behave similarly to their real counterparts when observed in Unity?
\item To what extent does it have population curves consistent with predictions from other ecosystem models, for example, Lotka-Volterra curves?
\item To what extent does the model make predictions that are consistent with common sense? For example, will the populations increase or decrease following certain changes?
\item To what extent is the model sensitive to randomness, small variations in the terrain, and parameter settings? 
Ecosystems seem to be subject to the \textit{butterfly effect} \cite{lorenz1963deterministic}: small changes in the initial conditions might create distinctly different outcomes and could make a difference between survival and eradication of certain populations. In fact, ecosystems depend on weather, which is subject to the butterfly effect.
\item To what extent does the model fit biological data collected from the real ecosystem under consideration? To what extent do real population data fit the predictions of the model?
\end{enumerate}

So far, we have worked with the first four criteria, but not the fifth. 
Case studies where model predictions are compared to biological data might increase confidence in the model, but never validate it once and for all. 
A shortcut might be to compare \ecotwin to other models that have in turn been validated against biological data to various degrees.

Clearly, one cannot have too high expectations on the realism of ecosystem models or the possibility of validating them. To begin with, our understanding of biology is not enough for making accurate models of animal behavior in complex ecosystems. Also, one needs to choose a reasonable level of detail for the model, which depends on the intended application. With \ecotwin we primarily aim for a model that makes reasonable predictions at the level of population dynamics. 
Quoting from \cite{trond2022contours}:
\begin{quote}
At their best, computational models suggest how things could \textit{possibly} work; fitted to data they might also propose how things could \textit{reasonably} work. Alas, asking if they can answer how things \textit{actually} are, is usually demanding too much.
\end{quote}

\subsection{Scalability}
\label{sec:scalability}

\ecotwin builds terrain models automatically from geographic data. 
To add new species of animals and plants, one may either use physical models that are available from Unity or define new ones. The physical design of the animals and plants plays no role in the \ecotwin simulations, except when visualizing the ecosystem development. 
What other properties need to be specified when defining new animals and plants, depends on the desired level of detail of the model.

To introduce new plants, one must specify where they grow and how they reproduce. Initially the plants might be randomly distributed according to a probability distribution with one probability for each land cover type. 
Plants in \ecotwin may reproduce by 
(i) regrowing on the same location (like our dandelions), 
(ii) by respawning on new random locations, 
(iii) by spreading via the roots to neighboring locations (like grass in \cite{glimmerfors2021combining}), or 
(iv) by spreading via the air (like dandelions in \cite{glimmerfors2021combining}). 
Plant models may die either from old age or from being consumed by animal models.

To introduce new animal species, one thing that needs to be specified is how they reproduce. For example, one may use the simple asexual reproduction mechanism used in this paper, or a detailed model of sexual reproduction with explicit representation of the genome. 
One must also specify the metabolism: the resource loss associated with different actions (and terrain), and the resource gain from eating various objects (plants and animals). 
One must also specify the perception and the utility function. 
Finally the new animal must be trained with DRL in a sequence of environments, including digital models of real terrain with all relevant plants and animals present. 

In summary, many things need to be specified, for each new animal species. This work is relatively straight-forward, however. The big gain is that we do not need to manually specify the animal's behavior. If we specify the reward system, then much of the rest can be left to RL algorithms. 
Regarding quantitative scalability, we have not stress tested Ecotwin, but we have seen that it runs well on ecosystem models with about 5000 re-spawning plants and 1000 animals on an ordinary laptop. 

\subsection{Relevance }
Economic activity is associated with a continuous stream of financial decisions, business decisions, political decisions, and engineering decisions. Those decisions should ideally take the interest of all stakeholders into account and consider aspects such as profitability, technical feasibility, and sustainability.  
Laws, industry regulations, business policies, procurement instructions, and business contracts frequently compel decision-makers to analyze the ecological consequences of the economic activity under consideration and produce an environmental impact assessment as a part of that process. 
This work can be done by experts from the private or public sector. 
A common practice is to not involve any ecological models at all in these processes. 
Part of the reason for that might be that there are no adequate ecological models within reach in many cases.
One may hope that the future will bring ecosystem models that are both easy to use and have enough predictive power to make them useful. 
This could contribute to improved decision-making on a broad scale and to more sustainable economic activities.
At present, the main challenge might be to introduce an ecosystem model that is clearly better than the default alternative of using no ecosystem model at all.

\section{Conclusion}
\label{sec:conclusion}

We have studied the feasibility, flexibility, and modeling power of the ecosystem simulator Ecotwin, which combines digital models of real terrain with animal models controlled by DRL. 
We showed how models of this kind might:
\begin{itemize}
\item contribute to theoretical biology, e.g., by shedding light on the role of the terrain and the ability to learn
\item analyze the local ecological consequences of a broad range of economic activities  
\item contribute to more sustainable decision-making in the private sector and policy-making in the public sector. 
\end{itemize}

Our plans for the future include developing a larger collection of animal models that can be inserted into arbitrary \ecotwin models, much like Lego pieces.
Moreover, we want to embark on the process of testing, improving, and validating \ecotwin by comparing its predictions to real biological data.


Digital twins of cars, ships, airplanes, spacecrafts, buildings, cities, landscapes, and transportation networks are widely used today for learning, understanding, testing, and predicting purposes \cite{jones2020characterising}. 
With terrain models based on geographic data and behavioral models based on AI, the time might have come also for digital twins of real ecosystems.

\bibliographystyle{splncs03_unsrt.bst} 
\bibliography{animatreferences,cvbibliography}




    



\end{document}